# Selective spin wave excitation in ferromagnetic (Ga,Mn)As layers by picosecond strain pulses

M. Bombeck[1], A.S. Salasyuk[1,2], B. A. Glavin[3], A.V. Scherbakov[2], C. Brüggemann[1], D.R. Yakovlev[1,2], V.F. Sapega[2], X. Liu[4], J. K. Furdyna[4], A.V. Akimov[2,5], and M. Bayer[1,2]

[1]*Experimentelle Physik 2, Technische Universität Dortmund, D-44227 Dortmund, Germany*

[2]*Ioffe Physical-Technical Institute, Russian Academy of Sciences, 194021 St. Petersburg, Russia*

[3]*Department of Theoretical Physics, V.E. Lashkaryov Institute of Semiconductor Physics, 03028 Kyiv, Ukraine*

[4]*Department of Physics, University of Notre Dame, Notre Dame, Indiana 46556, USA*

[5]*School of Physics and Astronomy, University of Nottingham, Nottingham NG7 2RD, UK*

ABSTRACT

We demonstrate selective excitation of a spin wave mode in a ferromagnetic (Ga,Mn)As film by picosecond strain pulses. For a certain range of magnetic fields applied in the layer plane only a single frequency is detected for the magnetization precession. We explain this selectivity of spin mode excitation by the necessity of spatial matching of magnon and phonon eigenfunctions, which represents a selection rule analogous to momentum conservation for magnon-phonon interaction in bulk ferromagnetic materials.

The enormous success of semiconductors is based on the possibility to tailor their electrical and optical properties almost arbitrarily. This has stimulated activities to seek such a level of control also for their magnetic properties which might allow all-in-one-chip solutions in information technology. Ferromagnetic semiconductors, on whose basis ultrafast magneto-electronic and magneto-optical devices [1] may become operational, are a key building block on this route. While great progress has already been achieved in this area, considerable obstacles still need to be overcome. This concerns not only the development of highest quality material, but also novel tools, distinctly different from established ones, for manipulating and controlling the magnetization may be required.

An example of a novel concept is the recent demonstration that the interaction between spin waves (SWs) may be used for spin current control [2]. During the last decade the underlying magnon excitations with frequencies in the GHz range have been intensively studied experimentally and theoretically in ferromagnetic (Ga,Mn)As [3-8]. These activities were focused on thin ferromagnetic films, in which SWs have a discrete frequency spectrum determined by parameters such as magneto-crystalline anisotropy, spin exchange interaction, layer thickness and boundary conditions. For application of SWs it is essential to have a technique which allows excitation of a single SW mode with particular frequency/wavelength. The most common technique to achieve this is microwave excitation at a frequency resonant with the SW mode [2-4,6-8]. While nicely functioning, microwave manipulation is limited to nanosecond time scales and cannot be scaled down to submicrometer spatial dimensions. Optical excitation of SWs has also been demonstrated, but this excitation is not frequency-selective [5].

In this letter we demonstrate the feasibility of a novel approach to excitation of a particular SW mode in a ferromagnetic layer by a picosecond strain pulse with a broad acoustic phonon spectrum that overlaps the SW frequencies. In contrast to the traditional methods, in the present work the selectivity of SW mode excitation is controlled by the spatial overlap of the SW mode and the component in the acoustic wavepacket spectrum with the same frequency as the SW. Usually, SWs have weak dispersion so that the frequency splitting between neighboring localized SW modes is small, even though they have very different shapes. This could result in a more efficient selective excitation than based on the SW mode parity of frequency. However, the spatial SW mode structure clearly depends crucially on the magnetic boundary conditions, and thus the spin properties at the interfaces determine in many respects the possibility to achieve a situation where only a single SW mode is excited. The main aim of the present work is to show that despite of all potential complication reliable conditions for single SW mode excitation can be obtained in experiment.

The studied sample is a single $Ga_{0.95}Mn_{0.05}As$ layer with thickness $d$=200 nm grown by low-temperature molecular-beam epitaxy on top of a semi-insulating (001) GaAs substrate. The Curie temperature of the ferromagnetic layer is 60 K, and the saturation magnetization is 20 emu/cm$^3$. The layer is compressively strained normal to the growth direction leading to an in-plane orientation of the easy axis of magnetization.

The experiments are carried out at temperature $T$=6 K in a cryostat with a superconducting magnet. The external magnetic field **B** was applied in the layer plane parallel to the easy axis, denoted as $x$-axis. For picosecond strain pulse generation [9], optical pulses from a femtosecond laser (wavelength 800 nm, pulse duration 150 fs, pulse energy density 2 mJ/cm$^2$, repetition rate 100 kHz) excite a 100-nm thick Al film deposited on the back side of the GaAs substrate. The strain pulse injected into the GaAs substrate has a spatial extent of ~100 nm with a maximum amplitude ~1×10$^{-4}$. The pulse propagates normal to the layer through the GaAs substrate (along the $z$-axis) at longitudinal sound velocity $s$=4.8 km/s, and after $t_0$=$l_0$/$s$ ≈ 22 ns ($l_0$=105 μm is the GaAs substrate thickness) reaches the (Ga,Mn)As magnetic layer. There it passes through the layer, becomes reflected at the open sample surface with a π-phase shift, and travels back towards the GaAs substrate.

While propagating through the (Ga,Mn)As layer the strain pulse modifies the magneto-crystalline anisotropy at each spatial (*z*) position as function of time (*t*), causing the magnetization **M** to be turned out of the equilibrium orientation, which is approximately parallel to the [100] axis. The magnitude of this turn and the corresponding tilt angle can be found in earlier works [10]. After the strain pulse has left the magnetic film, the subsequent dynamics of **M**(*z,t*) is given by harmonic oscillations of the components $M_z$ and $M_y$.

In the experiments the effect on the magnetization dynamics induced by the strain pulse is measured by monitoring the Kerr rotation angle $\varphi(t)$ as function of delay of a probe laser pulse relative to the pump pulse, both taken from the same laser. The pump beam was additionally modulated by a mechanical chopper, and $\Delta\varphi(t) = \varphi(t) - \varphi_0$ was recorded ($\varphi_0$ being the Kerr rotation angle without strain pulses. Examples of the signals $\Delta\varphi(t)$ measured at *B*=100 and 250 mT for three polarization settings of the probe beam are shown in Fig.1 (a) and 1 (b). Different time regimes are seen for the oscillatory behavior of $\Delta\varphi(t)$. The high frequency features in the interval *t*<0.1 ns results mainly from elasto-optical effects, which are described in detail in the recent work by Thevenard et al. [11]. For times *t*>0.1 ns, i.e. after the strain pulse has left the (Ga,Mn)As layer, the oscillations correspond exclusively to the magnetization precession. In the experimentally used geometry for which the direction of **M**$_0$ is close to the [100] axis the evolution of $\Delta\varphi(t)$ depends on the angle between the probe polarization plane and the [100] direction, $\psi$, and may be written as:

$$\Delta\varphi(t) = a\overline{M}_z(t) + b\overline{M}_y(t)\cos 2\psi + c(t)\sin 2\psi. \tag{1}$$

The first two terms in Eq. (1) reflect the dynamics of the layer magnetization, which in general is spatially non-uniform. Here, *a* and *b* are constants, and $\overline{M}_{z,y}(t)$ are determined by spatial distributions of the magnetization and the probe light field inside the magnetic layer. In case of no light absorption and no reflection at the (Ga,Mn)As/GaAs interface we have

$$\overline{M}_z(t) = \frac{1}{d}\int_0^d M_z(z,t)\cos[2k_{ph}(z-d)]dz \quad \text{and} \quad \overline{M}_y(t) = \frac{1}{d}\int_0^d M_y(z,t)\sin[2k_{ph}(z-d)]dz,$$ where the layer is located at $0 < z < d$ and $k_{ph}$ is the photon wavenumber of the probe light in the magnetic layer. The contributions of the $z$ and $y$ magnetization components to the rotation of probe polarization are due to the magneto-optical anisotropy and are governed by the circular [12] and giant linear [13] dichroism in (Ga,Mn)As, respectively. The third term in Eq. (1) describes the dynamical photoelastic perturbation induced by the strain pulse in the presence of a static magneto-optical anisotropy of the magnetic layer [11]. The constants $a$, $b$, and the dependence $c(t)$ are not known precisely. Nevertheless, Eq.(1) allows us to extract $\overline{M}_z(t)$ and $\overline{M}_y(t)$ from measurements of $\Delta\varphi(t)$ for three different probe beam polarizations **e**: e.g., for **e**∥**M₀**; **e**⊥**M₀**; and **e**∠**M₀** with an angle of 45 degrees between **e** and **M₀**. The resulting evolutions of $\overline{M}_z(t)$ and $\overline{M}_y(t)$ for the two values of $B$ are shown in Fig. 1 (c) and 1 (d).

Figure 2 (a) shows fast Fourier transform spectra of $\overline{M}_z(t)$ for different $B$. The spectra obtained from $\overline{M}_y(t)$ look similar. Generally, two spectral lines are seen whose low and high central frequencies, $f_l$ and $f_h$, shift smoothly to higher values with increasing $B$, while the spacing $\Delta f = f_h - f_l \approx 2$ GHz between them remains almost constant, see Fig. 2(b). The amplitudes of the two spectral lines vary with magnetic field. The most interesting feature in this respect is, that at fields around $B=B_0=225\pm25$ mT only a single line corresponding to the lower frequency component is observed. This is also demonstrated in Fig. 2 (c) which shows the peak intensities of the low and high frequency spectral lines versus $B$. Both vary non-monotonically with $B$, and the high frequency spectral line disappears around $B=B_0$, while the low frequency one has a maximum at these fields.

The dependence of the spectrum on magnetic field is the key experimental observation of the present work, and for its explanation we use the concept of standing SWs in the (Ga,Mn)As layer. Earlier work [5], in which SWs were excited optically, has also demonstrated doublet of lines in the SW spectrum, but the amplitudes of the peaks did not depend on $B$. Thus, the present tool using picosecond strain pulses represents a unique instrument for controlled selective SW excitation.

The physics underlying this SW excitation are related to the spin-phonon interaction in ferromagnetic materials, as discussed in literature [14]. In bulk materials energy and momentum conservation for spin-phonon interaction result in strict selection rules for the SW excitation or, in the case of strong coupling, for excitation of hybrid magnon-phonon modes. For thin films momentum conservation is relaxed so that a monochromatic acoustic wave may excite a resonant standing SW independent of its wavelength [15]. In our experiments the strain pulse corresponds to an acoustic phonon wavepacket that contains a broad distribution of frequencies, so that excitation cannot be considered as monochromatic. Nevertheless, we show below that an ultrashort strain pulse, propagating through the layer in forward and subsequently in backward direction, excites SW modes whose amplitudes are strongly dependent on the SW frequency and thus on $B$. Moreover for certain conditions only a single SW mode may be excited by the strain pulse. For this purpose we analyze the magnetization dynamics by the Landau-Lifshitz equation [16]:

$$\frac{\partial \mathbf{M}}{\partial t} = -\gamma \mathbf{M} \times \left( -\nabla_{\mathbf{M}} F + \frac{A}{M_0} \nabla^2 \mathbf{M} \right), \qquad (2)$$

where $\gamma$, $F$, $A$ and $M_0$, are the gyromagnetic ratio, the magnetic free energy density, the spin stiffness constant, and the magnetization magnitude, respectively. $F$ contains contributions determined by the magnetic layer properties and the applied magnetic field. In addition to the case of an unstrained crystal, $F$ contains magneto-elastic terms for biaxially strained (Ga,Mn)As [17].

The strain pulse propagating through the magnetic layer causes a variation of $F$ in time, $t$, and space, $z$, resulting in a complicated trajectory of **M.** When the strain pulse leaves the magnetic layer, **M** continues to precess around its equilibrium position while relaxing towards **M₀** [note, that relaxation is not included in Eq. (2)]. During this precession the asymptotic solutions of the magnetization components $M_i$ can be obtained in linear approximation in which the deviation from steady-state $\delta \mathbf{M}(z,t) = \mathbf{M}(z,t) - \mathbf{M}_0$ is written as superposition of standing SW eigenmodes $S_i^{(n)}(z)$:

$$\delta M_i(z,t) = \sum_{n=0}^{\infty} C^{(n)} S_i^{(n)}(z) \cos(\omega_n t + \phi_n), \qquad (3)$$

where the $C^{(n)}$ and $\phi_n$ are the stationary amplitude and phase of the $n$-th mode with frequency $\omega_n$ ($n$=0, 1, 2…), respectively. Calculation of the mode amplitudes $C^{(n)}$ shows that they are proportional to the overlap integral $\int_0^d S_i^{(n)}(z)\varepsilon_n(z)dz$, where $\varepsilon_n(z)$ is the Fourier component in the strain pulse with phonon frequency identical to that of the SW $\omega_n$. Assuming 100% reflectivity of the strain pulse at the sample surface one finds that $\varepsilon_n(z) \sim \sin[\omega_n(z-d)/s]$.

The SW eigenmodes $S_i^{(n)}(z)$ are controlled by the magnetic boundary conditions for solving Eq.(2). One may assume pinning of the magnetization at the interfaces, i.e. $\delta \mathbf{M}(0,t) = \delta \mathbf{M}(d,t) = 0$. Then, by ignoring volume inhomogeneities of the magnetic properties in the film, one easily obtains $S_i^{(n)} \sim \sin[\pi(n+1)z/d]$ so that excitation the $n$-th SW mode is possible only if:

$$\frac{\omega_n}{s} = \frac{\pi}{d}(n+1). \qquad (4)$$

This condition corresponds to the case in which the magnetization vector $\mathbf{M}$ at $z\approx 0$ performs exactly ($n$+1) full revolutions while the strain pulse is traveling forward and backward through the magnetic layer.

Within the macroscopic Landau-Lifshitz approach, the boundary conditions can be introduced through the surface magnetic energy $F_{surf}$. Most commonly, it is assumed that $F_{surf} = K_s \cos^2\theta$ (where $\theta$ is the angle between $\mathbf{M}$ at the surface and the normal to the surface $\mathbf{n}$ and $K_s$ is the surface magnetic energy parameter). This corresponds to the surface easy axis being either normal ($K_s < 0$) or parallel to the surface ($K_s > 0$). Then the boundary conditions can be written as $\mathbf{M} \times \left( A \frac{\partial \mathbf{M}}{\partial \mathbf{n}} + 2\mathbf{n}K_s \cos\theta \right) = 0$ [18]. It is possible to show that for large positive $K_s$ value the SW modes are very similar to those obtained for the pinning conditions.

The results of calculations using the pinning boundary conditions are shown in Fig. 3. The solid lines in Fig. 3(a) show the spatial shapes of $S_z^{(n)}(z)$ for the three lowest modes ($n$=0, 1 and 2) and the dashed line gives $\varepsilon_0(z)$ for the value of $B$ at which the condition Eq. (4) is met for the fundamental mode ($n=0$). Fig. 3 (a) demonstrates clearly that for these resonant conditions only the overlap integral for the SW mode with $n$=0 is non-zero. Therefore, only this SW mode is excited selectively by the picosecond strain pulse. The mode amplitudes $C^{(n)}$ as functions of the dimensionless parameter $\omega_n d/(\pi s)$ are shown in Fig.3 (b) for $n$=0, 1 and 2. The vertical dashed line in Fig. 3(b) indicates the value $\omega_n d/(\pi s)$ =1 for which Eq.(2) is fulfilled for $n$=0 and therefore only one mode ($n$=0) has non-zero amplitude in this case and thus becomes exclusively excited.

Remarkably, the single line observed in the experimental spectrum around $B=B_0$ [see Fig.2 (a)] has the frequency $f_l$=12 GHz, that corresponds to the fundamental radial frequency $\omega_0$ given by Eq. (4). Thus for the assumed spin pinning boundary conditions we get excellent agreement with experiment if we attribute the lower frequency ($f_l$) spectral line with the fundamental SW mode ($n$=0). However, we can not unambiguously attribute the high-frequency ($f_h$) spectral line in the experiment to the SW mode with $n$=1. The theoretically calculated frequency dependences of $C^{(n)}$ for $n>0$ look quite similar to each other and, for our experimental conditions, where the probe wavelength is close to the fundamental absorption band, we cannot analyze quantitatively the efficiency of optical detection for different SW modes. Thus it is difficult to associate the high frequency spectral line to a certain SW mode with $n$>0. However, this does not affect our unique conclusion about the selective excitation of the fundamental SW mode with $n$=0 at $B=B_0$.

It is worth mentioning that although a perfect selectivity in excitation requires spin pinning at the film interfaces, a high degree of selectivity is expected to be attainable also for a wider class of boundary conditions, especially for the fundamental mode $n=0$. This should happen, in particular, if the mode's structure is influenced by volume inhomogeneities of the magnetic anisotropy parameters, suppressing the magnitude of magnetization near the interfaces [6]. Although it is difficult to analyze the mode amplitudes for this case quantitatively, the fast spatial oscillations of the eigenfunctions $S_z^{(n)}(z)$ with $n>0$ suggest that the efficiency of their excitation is small if the resonance condition (4) is fulfilled for the ground mode.

In summary, we have demonstrated that a single spin wave mode in a (Ga,Mn)As layer can be excited selectively by a picosecond strain pulse. Experimentally we observe a strong dependence of the excited SW spectrum on the magnetic field and, consequently, on the SW frequency. Only one spectral line, attributed to the fundamental SW mode, is observed when the magnetization precession period is equal to the strain pulse travel time forward and backward through the magnetic layer. The observed selectivity is attributed to the excitation efficiency, which depends crucially on the spatial shape of the magnetization of distinct SW modes and their overlap with the corresponding spatial Fourier component in the acoustic wavepacket − representing a kind of selection rule. Such selective SW excitation by picosecond acoustic techniques is a prospective tool for spin current manipulations in devices in which hypersonic nanostructures, like phonon cavities [19] or sasers [20], are combined with electro-magnetic and opto-magnetic components in a single semiconductor chip. The understanding established here on the selectivity of SW excitation gives a useful guide to tailor both magnetic layer and phonon pulse such that SWs of particular frequencies can be excited.

The work was supported by the Deutsche Forschungsgemeinschaft (BA 1549/14-1), the Russian Foundation for Basic Research, the Russian Academy of Sciences, the European

Community's Seventh Framework Programme (grant n$^o$ 214954), and the National Science Foundation Grant DMR10-05851.

**Figure captions.**

**Figure 1.**

Upper panels: Kerr rotation signals measured for various polarizations of the probe beam at applied magnetic fields $B$=100 mT (a) and 250 mT (b); the horizontal bars indicate the time intervals during which the strain pulse is present inside the film. Lower panels: Temporal evolutions of mean magnetization projections $M_z(t)$ and $M_y(t)$ for the same $B$ as in (a), obtained using Eq. (1) from the measured Kerr rotation signals.

**Figure 2.**

(a) Amplitude spectra for the $z$-projection of the magnetization for different applied magnetic fields, indicated at each curve. The $B$ values for which selective excitation of the low energy SW mode is detected are highlighted. (b) Central frequencies of the excited SW modes as functions of $B$; (c) Magnetic field dependencies of the peak intensities of the low- and high-frequency SW modes. The arrows in (b) and (c) indicate the frequency and magnetic field around which when the selective excitation is observed.

**Figure 3.**

(a) Spin wave eigenfunctions (solid lines) of the three lowest modes. The dotted line shows the spatial Fourier component of the acoustic wave packet in the strain pulse which corresponds to the phonon frequency for which the period of one full magnetization revolution is equal to the propagation time of the strain pulse forward and backward through the magnetic film [see Eq. (4) for $n$=0]. (b) Dependencies of the SW mode amplitude on the normalized resonance frequency for the three lowest modes. The vertical dashed-dotted line corresponds to the frequency at which the ground mode $n=0$ is excited selectively. The vertical arrows indicate the frequencies related to the experimental conditions at $B=0$ and $B=500$ mT.

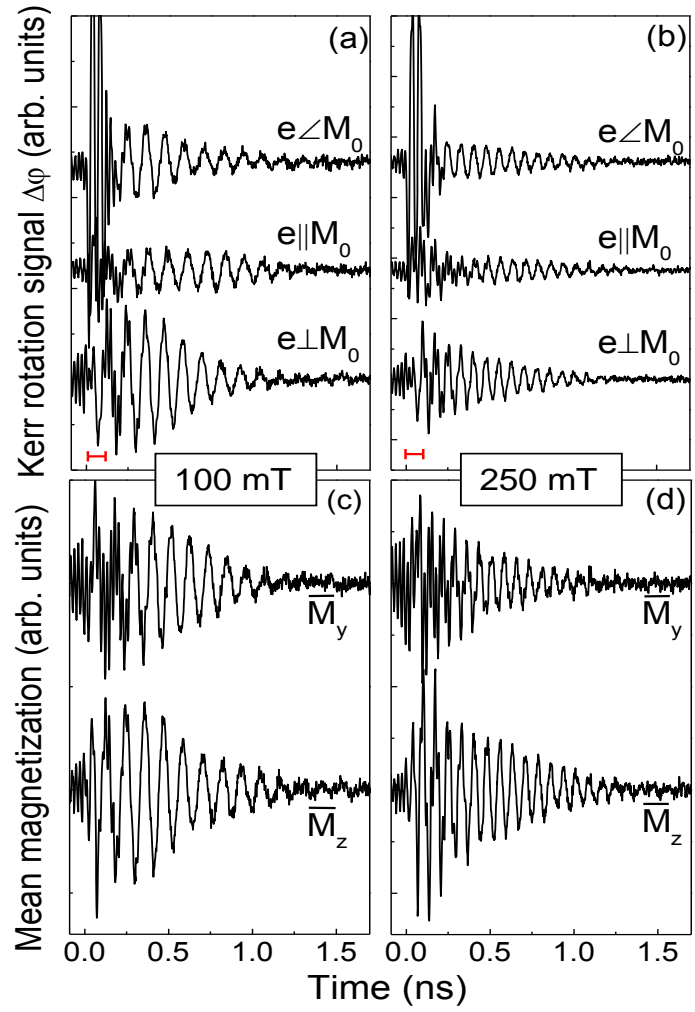

Figure 1. *M. Bombeck et al. "Selective spin wave excitation…"*

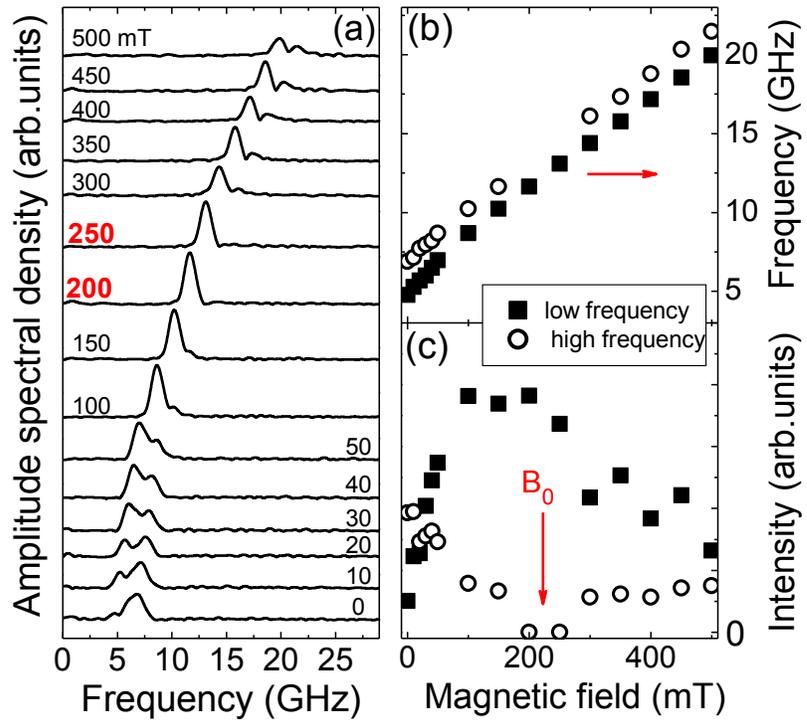

Figure 2. *M. Bombeck et al. "Selective spin wave excitation…"*

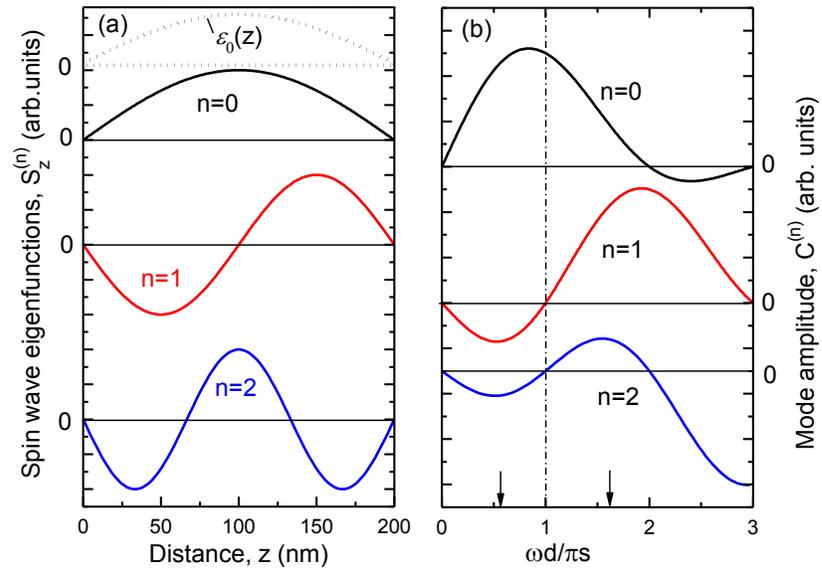

Figure 3. *M. Bombeck et al. "Selective spin wave excitation…"*